\documentclass[twocolumn,showpacs,preprintnumbers,amsmath,amssymb]{revtex4}

\usepackage{graphicx}
\usepackage{dcolumn}
\usepackage{bm}

\begin{document}

\preprint{}
\input{epsf.tex}

\epsfverbosetrue

\title{Optical Properties of Collective Excitations for Finite Chains of Trapped Atoms}

\author{Hashem Zoubi, and Helmut Ritsch}

\affiliation{Institut fur Theoretische Physik, Universitat Innsbruck, Technikerstrasse 25, A-6020 Innsbruck, Austria}

\date{24 June, 2011}

\begin{abstract}
Resonant dipole-dipole interaction modifies the energy and decay rate of electronic excitations for finite one dimensional chains of ultracold atoms in an optical lattice. We show that collective excited states of the atomic chain can be divided into dark and bright modes, where a superradiant mode with an enhanced collective effective dipole dominates the optical scattering. Studying the generic case of two chain segments of different length and position exhibits an interaction blockade and spatially structured light emission. Ultimately, an extended system of several interfering segments models a long chain with randomly distributed defects of vacant sites. The corresponding emission pattern provides a sensitive tool to study structural and dynamical properties of the system.
\end{abstract}

\pacs{37.10.Jk, 42.50.-p, 71.35.-y}

\maketitle

\section{Introduction}

Trapping and manipulating ultracold atoms in an array of potentials created a turning point in the field of quantum fluids \cite{Metcalf} and their application for quantum information processing \cite{Zeilinger}. Implementing a Mott insulator of strongly interacting bosons in an optical lattice laid the foundations to study and simulate many strongly correlated quantum many-body phases \cite{Dalibard}, with a big impact on condensed matter physics \cite{Lewenstein}. Nowadays many types of tunable quantum many-body systems are implemented to the study of central condensed matter phenomena, as quantum phase transitions, superfluidity and quantum magnetisms. Counter propagating laser beams produce a designable optical lattices where atoms can hop among nearest sites with tunable on-site interactions, where in controlling the laser intensity the system casts from the superfluid into the Mott insulator phase \cite{Jaksch}. This quantum phase transition is predicted by the Bose-Hubbard model and realized experimentally \cite{Bloch}, where in the Mott insulator the ultracold atoms are perfectly arranged on the lattice with fixed number of atoms per site.

In any practical setup, however, the formation process of a Mott insulator is accompanied by the appearance of defects \cite{Bloch,Spielman}. For example, a non ideal Mott insulator of filling factor one, namely of single atom per site, contains many vacancies that distributed randomly on the lattice. Such vacancies are unavoidable and must be taken into account in any quantitatively precise treatment of the Mott insulator. The vacancies are mobile and hop from site to site with the same parameter as the atom hopping parameter among nearest neighbor lattice sites. For experiments concerned with electronic excitations and optical processes in optical lattices, the light-matter coupling and electrostatic interaction parameters are usually much larger than the atom hopping parameter, and hence the vacancies can taken to be frozen on the lattice.

A different approach to form one dimensional optical lattices has been opened in the last years which rests on optical fibers \cite{Rauschenbeutel,Nayak}. For tapered optical nanofibers the strong evanescent field surrounding the fiber is efficiently used for trapping, manipulating and detecting the atoms \cite{Vetsch}. The guided modes of ultrathin optical fibers, with diameters smaller than the wavelength of the guided light, exhibit strong transverse confinement and pronounced evanescent field. Interference of two color evanescent fields surrounding an optical nanofiber give rise to an array of optical microtraps. The system was realized recently for cesium atoms interacting with a multicolor evanescent field surrounding an optical nanofiber, where the atoms are localized in a one dimensional lattice parallel to the nanofiber \cite{Vetsch}. Due to the small trapping volumes the occupancy of each site is at most one and the filling factor is half.

In another approach the high degree of control of experimental parameters in optical lattices has been demonstrated to be implemented to the level of a single atom at a specific lattice site, the fact that should revolutionized the field of optical lattice ultracold atoms and its application to condensed matter physics \cite{Sherson,Bakr}. Furthermore, the flip of a spin (change of the internal hyperfine state) of individual atoms of certain lattice sites in the Mott insulator phase can performed in a controlled way by the help of a tightly focused laser beam. Such a technique provides a tool to create arbitrary spin patterns in a two dimensional array of perfectly arranged atoms in the Mott insulator phase. Several segments of one dimensional finite arrays with different lengths and shapes are realized \cite{Weitenberg}.

The similarity between optical lattice ultracold atoms in the Mott insulator phase and molecular crystals encourages us to introduce collective electronic excitations into such a system \cite{ZoubiA}. The physics is similar to Frenkel excitons in which an electronic excitation can be transferred among atoms at different sites and delocalized in the lattice due to electrostatic interactions, i.e. resonance dipole-dipole interactions \cite{Davydov}. In exploiting the lattice symmetry, an electronic excitation can be represented as a wave that propagates in the lattice, which is a quasi-particle termed an exciton \cite{Agranovich}. We extensively studied excitons in one and two dimensional optical lattices \cite{ZoubiA,ZoubiB}, and we examined their life times and emission patterns into free space \cite{ZoubiC,ZoubiD}, where we found that excitons can be metastable or superradiant. So far we treated large one and two dimensional optical lattices, where excitons behave as propagating plane waves. The inclusion of defects in two dimensional optical lattices investigated by us in \cite{ZoubiE} and gives the scattering of excitons off such defects.

In the present paper we investigate few ultracold atoms in one dimensional optical lattices. Motivated by the recent experiments in the quantum engineering of many-particle systems to form finite chains of single atom sites in planar optical lattices \cite{Weitenberg}, we study a system of several one dimensional finite chains. In our previous work \cite{ZoubiF} we considered an infinite ideal optical lattice with filling factor one formed by tapered optical nanofiber, while in the presently realized system the filling factor is half \cite{Vetsch}. Here we treat a finite one dimensional optical lattice that contains many vacancies which distributed randomly along the lattice.

We start by discussing an electronic excitation in finite one dimensional optical chain with one atom per site in exploiting the formation of collective electronic excitations induced by resonance dipole-dipole interactions and in using the lattice symmetry. Due to the finite length of the system, the collective states are standing waves with dark and bright modes, where the first bright mode found to be superradiant. For each of the bright collective modes we calculate the collective transition dipole, damping rate and emission pattern into free space. Next, we investigate the case of several finite one dimensional optical lattices of different lengths, where we examine the emitted light of the whole system. The atomic segments are located along the same one dimensional optical lattice and are separated by vacancies of empty sites. We concentrate in the case of two finite atomic segments, where coherence effects and quantum beats are emphasized.

The paper is organized as follows. In section 2 we introduce collective electronic excitations for finite one dimensional optical lattice, where we calculate their damping rate and emission pattern. The finite one dimensional optical lattice with defects of vacancies is treated in section 3. Section 4 is mainly about two segments of finite optical lattices where we present the results for a specific case. A summary is included in section 5.

\section{Electronic Excitations in Finite Optical Lattices}

We consider a finite one dimensional optical lattice of ultracold atoms with one atom per site. The number of lattice sites is $N$ with lattice constant $a$, as seen in figure (1). The atoms are considered to be two-level systems with electronic transition energy of $E_A$. An electronic excitation can delocalize in the lattice due to resonance dipole-dipole interactions, namely an electronic excitation can transfer among the lattice sites \cite{ZoubiA}. The electronic excitation Hamiltonian is given by
\begin{equation}
H_{ex}=\sum_nE_A\ B_n^{\dagger}B_n+\sum_{nm}J\ B_n^{\dagger}B_m,
\end{equation}
where $B_n^{\dagger}$ and $B_n$ are the creation and annihilation operators of an excitation at site $n$. For a single excitation the operators obey boson commutation relations. Energy transfer is assumed only among nearest neighbor sites, with the coupling strength
\begin{equation} \label{Coupling}
J=\frac{\mu^2}{4\pi\epsilon_0a^3}\left(1-3\cos^2\theta\right),
\end{equation}
where $\mu$ is the magnitude of the electronic excitation transition dipole, which makes an angle $\theta$ with the lattice direction, see figure (1).

\begin{figure}[h!]
\centerline{\epsfxsize=8cm \epsfbox{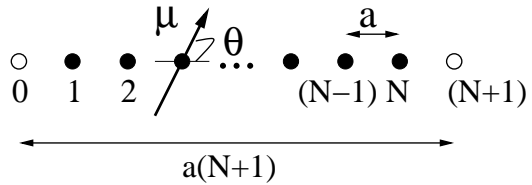}}
\caption{A finite optical lattice of $N$ sites with one atom per site. The lattice constant is $a$ and the lattice length is $L=a(N+1)$, where the edges empty sites are added to get fixed boundary condition. The transition dipole $\mbox{\boldmath$\mu$}$ makes an angle $\theta$ with the lattice direction.}
\end{figure}

The delocalization of an electronic excitation in the lattice give rise to a collective mode. The finite dimension of the lattice results in a standing wave collective excitation. In exploiting the lattice symmetry, the Hamiltonian is diagonalized by applying the transformation
\begin{equation} \label{Trans}
B_n=\sqrt{\frac{2}{N+1}}\sum_k\sin\left(\frac{\pi nk}{N+1}\right)\ B_k,
\end{equation}
where $k=1,\cdots,N$, as the number of the lattice sites. Here we added two additional empty sites at the two lattice edges, at $n=0$ and $n=N+1$, where the collective electronic excitation wave vanishes at these sites. These empty sites are defects in a larger optical lattice. The first four modes are plotted in figure (2) for the case of $N=10$. The diagonalization yields the collective electronic excitation Hamiltonian
\begin{equation}
H_{ex}=\sum_kE_k\ B_k^{\dagger}B_k,
\end{equation}
with the collective electronic excitation dispersion
\begin{equation}
E_k=E_A+2J\cos\left(\frac{\pi k}{N+1}\right).
\end{equation}
The single electronic transition energy splits into $N$ discrete collective electronic excitation energies. In figure (3) we plot the collective electronic excitation energy dispersion. We use the numbers: $E_A=1\ eV$, $a=1000\ A$, $\mu=1\ eA$ and $\theta=0$ with $N=10$.

\begin{figure}[h!]
\centerline{\epsfxsize=8cm \epsfbox{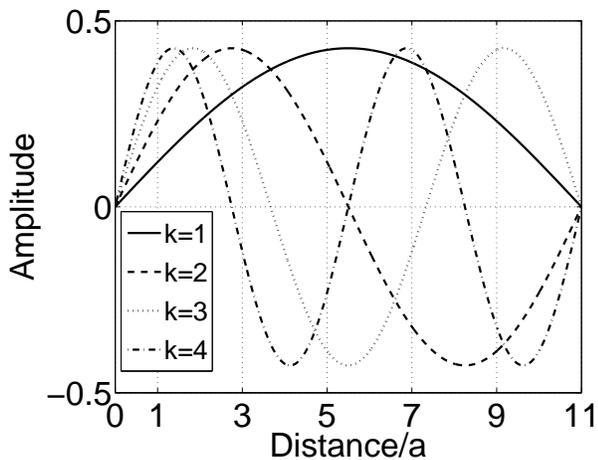}}
\caption{The collective electronic states. For $N=10$ we plot the first four modes. The odd modes are symmetric and the even modes are antisymmetric.}
\end{figure}

\begin{figure}[h!]
\centerline{\epsfxsize=8cm \epsfbox{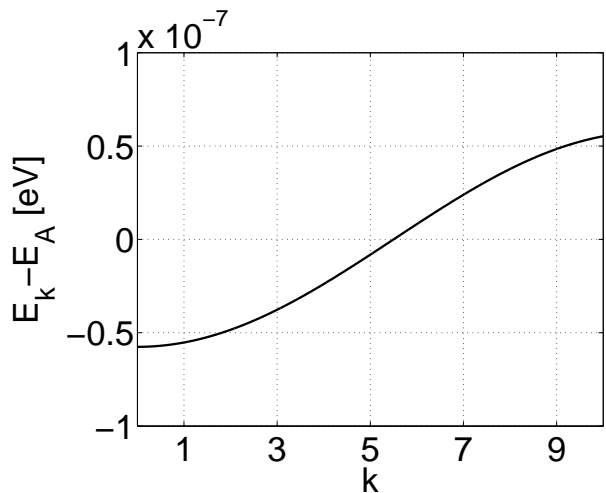}}
\caption{The energy dispersion for the collective electronic excitation, $E_k-E_A$ vs. $k$, for $N=10$ and $\theta=0$.}
\end{figure}

\subsection{Collective Transition Dipoles and Damping Rates}

The excitation transition dipole operator is
\begin{equation}
\hat{\mbox{\boldmath$\mu$}}=\mbox{\boldmath$\mu$}\sum_n\left(B_n+B_n^{\dagger}\right).
\end{equation}
Using the transformation (\ref{Trans}), we can write
\begin{equation}
\hat{\mbox{\boldmath$\mu$}}=\sum_k\mbox{\boldmath$\mu$}_k\left(B_k+B_k^{\dagger}\right),
\end{equation}
where the collective transition dipole of mode $k$ is
\begin{equation}
\mbox{\boldmath$\mu$}_k=\mbox{\boldmath$\mu$}\sqrt{\frac{2}{N+1}}\sum_n\sin\left(\frac{\pi kn}{N+1}\right).
\end{equation}
For odd $k$-s, $(k=1,3,\cdots)$, we get
\begin{equation}
\sum_n\sin\left(\frac{\pi kn}{N+1}\right)=\cot\left(\frac{\pi k}{2(N+1)}\right),\end{equation}
and for even $k$-s, $(k=2,4,\cdots)$, the summation vanishes. The collective transition dipole for $(k=1,3,\cdots)$ is
\begin{equation}
\mbox{\boldmath$\mu$}_k=\mbox{\boldmath$\mu$}\sqrt{\frac{2}{N+1}}\cot\left(\frac{\pi k}{2(N+1)}\right),
\end{equation}
and for $(k=2,4,\cdots)$ is $\mbox{\boldmath$\mu$}_k=0$. In figure (4) we plot the scaled transition dipole of the collective electronic excitations relative to a single excited atom, namely $|\mbox{\boldmath$\mu$}_k|/|\mbox{\boldmath$\mu$}|$. We conclude that the even modes are {\it dark}, and the odd modes are {\it bright}. Now we calculate the damping rate and emission pattern into free space for the bright modes.

\begin{figure}[h!]
\centerline{\epsfxsize=8cm \epsfbox{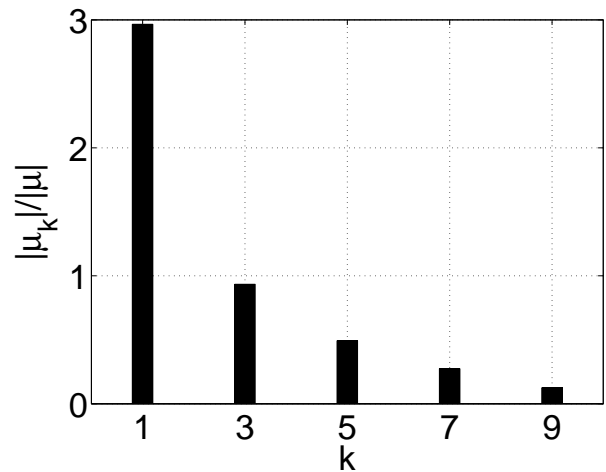}}
\caption{The collective transition dipole relative to a single excited atom, $|\mbox{\boldmath$\mu$}_k|/|\mbox{\boldmath$\mu$}|$ vs. $k$, for $N=10$.}
\end{figure}

The damping rate for a bright collective electronic excitation of mode $k$ is obtained from the Fermi Golden rule, which is similar to an excited atom damping rate but in replacing the excited atom transition dipole $\mbox{\boldmath$\mu$}$ with the collective transition dipole $\mbox{\boldmath$\mu$}_k$, which hold when the size of the chain is smaller than the wave length of the atomic transition. We have
\begin{equation}
\Gamma_k=\frac{\mu^2E_k^3}{3\pi\epsilon_0\hbar^4c^3}\left(\frac{2a}{L}\right)\cot^2\left(\frac{\pi ka}{2L}\right),
\end{equation}
where the lattice length is $L=a(N+1)$. The first mode $(k=1)$ without nodes, is found to be {\it superradiant}, with damping rate nine times larger than the second bright mode $(k=3)$. A significant deviation is obtained from that of a single excited atom damping rate, where for a single site we have $L=2a$ and $E_k=E_A$, we get the known result \cite{Loudon}
\begin{equation}
\Gamma_A=\frac{\mu^2E_A^3}{3\pi\epsilon_0\hbar^4c^3}.
\end{equation}
In figure (5) we plot the collective electronic excitation damping rate relative to a single excited atom, that is $\Gamma_k/\Gamma_A$, where we used the previous numbers.

\begin{figure}[h!]
\centerline{\epsfxsize=8cm \epsfbox{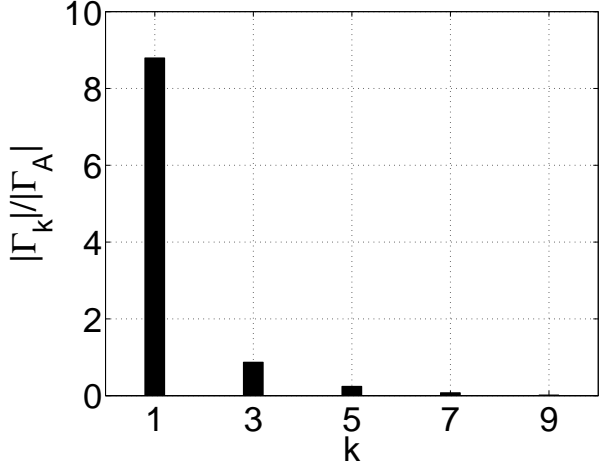}}
\caption{The collective electronic excitation damping rate relative to a single excited atom, that is $\Gamma_k/\Gamma_A$ vs. $k$, for $N=10$.}
\end{figure}

\subsection{Emission Pattern}

The emission pattern for a collective electronic excitation of mode $k$ into free space is obtained from that of an excited atom by using the collective mode transition dipole and energy. We calculate the positive electric field operator, in the far field zone at distance $|{\bf r}-{\bf R}|$ from the finite optical lattice, where ${\bf r}$ is the observation point and ${\bf R}$ is collective transition dipole average position, e.g. the center of the lattice, which is a good approximation in the limit of $|{\bf r}-{\bf R}|\gg L$. The result is given by
\begin{eqnarray}
\hat{\bf E}_k^{(+)}({\bf r},t)&=&i\frac{\mu E_k^2}{4\pi\epsilon_0\hbar^2c^2}\sqrt{\frac{2a}{L}}\cot\left(\frac{\pi ka}{2L}\right)\frac{\sin\phi}{|{\bf r}-{\bf R}|} \nonumber \\
&\times&B_k\left(t-\frac{|{\bf r}-{\bf R}|}{c}\right),
\end{eqnarray}
where $\phi$ is the angle between the collective transition dipole $\mbox{\boldmath$\mu$}_k$ and the distance vector ${\bf r}-{\bf R}$, as seen in figure (6). For the case of a single atom we get the result \cite{Loudon}
\begin{equation}
\hat{\bf E}_A^{(+)}({\bf r},t)=i\frac{\mu E_A^2}{4\pi\epsilon_0\hbar^2c^2}\frac{\sin\phi}{|{\bf r}-{\bf R}|}B\left(t-\frac{|{\bf r}-{\bf R}|}{c}\right).
\end{equation}

\begin{figure}[h!]
\centerline{\epsfxsize=5cm \epsfbox{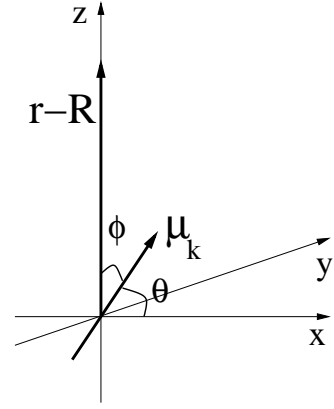}}
\caption{The collective transition dipole is located at ${\bf R}$, and the observation point at ${\bf r}$, where the vector $({\bf r}-{\bf R})$ makes an angle $\phi$ with $\mbox{\boldmath$\mu$}_k$.}
\end{figure}

The expectation value of the collective operators is \cite{Loudon}
\begin{eqnarray}
&&\langle B_k^{\dagger}(t-|{\bf r}-{\bf R}|/c)B_k(t-|{\bf r}-{\bf R}|/c)\rangle  \nonumber \\
&=&\langle B_k^{\dagger}(0)B_k(0)\rangle e^{-\Gamma_k(t-|{\bf r}-{\bf R}|/c)}.
\end{eqnarray}
The far field electric field intensity at the observation point ${\bf r}$ is defined by $I_k=\frac{1}{2}\epsilon_0c\langle\hat{\bf E}_k^{(-)}\hat{\bf E}_k^{(+)}\rangle$, where for a single collective excitation of mode $k$, that is $\langle B_k^{\dagger}(0)B_k(0)\rangle=1$, we get
\begin{eqnarray}
I_k&=&\frac{\mu^2 E_k^4}{16\pi^2\epsilon_0\hbar^4c^3}\left(\frac{a}{L}\right)\cot^2\left(\frac{\pi ka}{2L}\right)\frac{\sin^2\phi}{|{\bf r}-{\bf R}|^2} \nonumber \\
&\times& e^{-\Gamma_k(t-|{\bf r}-{\bf R}|/c)}.
\end{eqnarray}
For a single excited atom we have
\begin{equation}
I_A=\frac{\mu^2 E_A^4}{32\pi^2\epsilon_0\hbar^4c^3}\frac{\sin^2\phi}{|{\bf r}-{\bf R}|^2}e^{-\Gamma_A(t-|{\bf r}-{\bf R}|/c)}.
\end{equation}
The emission is dominated by the superradiant mode.

\section{Finite Opotical Lattices with Defects}

In the formation of one dimensional optical lattices with one atom per site many vacancies appear in the lattice. Lets assume that the vacancies are randomly distributed, such that in a long optical lattice we get many lattice segments of one, two, three, etc... atoms, as plotted in figure (7). We assume that the vacancies are localized where we neglect their hopping in the lattice. As we treat here electronic excitations in optical lattices where their hopping is much faster than the atom hopping we can consider the vacancies as frozen on the lattice. First, we treat each segment separately where we use the previous results concerning the formation of collective electronic excitations and their collective transition dipoles, damping rates and emission patterns. Furthermore, for simplicity we assume the lattice segments to be distributed in such a way that no two segment with the same number of sites are neighbors. Therefore we can neglect the hopping of a collective electronic excitations among different lattice segments as they are off resonance. We will discuss this point in much more details in the next section in treating two optical lattice segments. Next, for a given configuration of segments in a one dimensional optical lattice we calculate the emission pattern. We consider only the superradiant modes to be excited at each segment and which dominate the optical properties. Moreover we consider the case of at most a single electronic excitation at each segment. The lattice segments are assumed to be separated by a single empty site. These assumption simplify the calculation but they will not affect the physical conclusion. The only limitation is to stay far from the saturation regime, that is to be in the linear regime.

\begin{figure}[h!]
\centerline{\epsfxsize=8cm \epsfbox{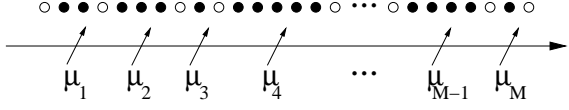}}
\caption{A finite optical lattice of $M$ segments of one atom per site that are separated by a single site vacancy. The collective transition dipoles of the superradiant modes are taken to be localized at the center of the segments.}
\end{figure}

We consider a one dimensional optical lattice with $M$ lattice segments. The $i$-th lattice segment contains $N_i$ sites and is of length $L_i=a(N_i+1)$. The optical lattice is along the $\hat{\bf x}$ axis. The lattice segment collective transition dipole of the superradiant mode is given by
\begin{equation}
\mbox{\boldmath$\mu$}_i=\mbox{\boldmath$\mu$}\sqrt{\frac{2a}{L_i}}\cot\left(\frac{\pi a}{2L_i}\right),
\end{equation}
with energy
\begin{equation}
E_i=E_A+2J\cos\left(\frac{\pi a}{L_i}\right).
\end{equation}
The transition dipoles are in the $(x-z)$ plane, with $\mbox{\boldmath$\mu$}=\mu(\cos\theta,0,\sin\theta)$, and the damping rate is
\begin{equation}
\Gamma_i=\frac{\mu^2E_i^3}{3\pi\epsilon_0\hbar^4c^3}\left(\frac{2a}{L_i}\right)\cot^2\left(\frac{\pi a}{2L_i}\right).
\end{equation}

The observation point ${\bf r}$ is taken to be in the $(x-z)$ plane. The center of the segment $i$ is at position ${\bf R}_i$, where ${\bf R}_i=(R_i,0,0)$. In the far zone field we assume that $|{\bf r}-{\bf R}_i|\ll L_i$, then we can assume the collective transition dipole to be localized at ${\bf R}_i$. The angle between the vector ${\bf r}-{\bf R}_i$ and the collective transition dipole $\mbox{\boldmath$\mu$}_i$ is $\phi_i$. The positive electric field operator of the collective transition dipole at segment $i$ at point ${\bf r}$ is
\begin{eqnarray}
\hat{\bf E}_i^{(+)}({\bf r},t)&=&i\frac{\mu E_i^2}{4\pi\epsilon_0\hbar^2c^2}\sqrt{\frac{2a}{L_i}}\cot\left(\frac{\pi a}{2L_i}\right)\frac{\sin\phi_i}{|{\bf r}-{\bf R}_i|} \nonumber \\
&\times&B_i\left(t-\frac{|{\bf r}-{\bf R}_i|}{c}\right)\hat{\bf e}_i,
\end{eqnarray}
where the direction of the electric field of the $i$-th segment at the observation point is given by the unit vector 
\begin{equation}
\hat{\bf e}_i=\hat{\bf y}\times\hat{\bf n}_i,
\end{equation}
where
\begin{equation}
\hat{\bf n}_i=\frac{{\bf r}-{\bf R}_i}{|{\bf r}-{\bf R}_i|}.
\end{equation}
For the collective atomic transition operators we can use the expectation values \cite{Loudon}
\begin{eqnarray}
&&\langle B_i(t-|{\bf r}-{\bf R}_i|/c)\rangle \nonumber \\
&=&\langle B_i(0)\rangle e^{-iE_i(t-|{\bf r}-{\bf R}_i|/c)/\hbar} e^{-\Gamma_i(t-|{\bf r}-{\bf R}_i|/c)/2}, \nonumber \\
&&\langle B_i^{\dagger}(t-|{\bf r}-{\bf R}_i|/c)B_i(t-|{\bf r}-{\bf R}_i|/c)\rangle \nonumber \\
&=&\langle B_i^{\dagger}(0)B_i(0)\rangle e^{-\Gamma_i(t-|{\bf r}-{\bf R}_i|/c)},
\end{eqnarray}
and
\begin{eqnarray}
&&\langle B_i^{\dagger}(t-|{\bf r}-{\bf R}_i|/c)B_j(t-|{\bf r}-{\bf R}_j|/c)\rangle \nonumber \\
&=&\langle B_i^{\dagger}(0)B_j(0)\rangle e^{-\Gamma_i(t-|{\bf r}-{\bf R}_i|/c)/2}e^{-\Gamma_j(t-|{\bf r}-{\bf R}_j|/c)/2} \nonumber \\
&\times&e^{iE_i(t-|{\bf r}-{\bf R}_i|/c)/\hbar}e^{-iE_j(t-|{\bf r}-{\bf R}_j|/c)/\hbar}.
\end{eqnarray}

The total electric field at the observation point is
\begin{equation}
\hat{\bf E}^{(+)}({\bf r},t)=\sum_i\hat{\bf E}_i^{(+)}({\bf r},t),
\end{equation}
and the intensity is
\begin{equation}
I({\bf r},t)=\frac{1}{2}\epsilon_0c\langle\hat{\bf E}^{(-)}({\bf r},t)\mbox{\boldmath$\cdot$}\hat{\bf E}^{(+)}({\bf r},t)\rangle.
\end{equation}
Explicitly we can write
\begin{equation}
I({\bf r},t)=\sum_iI_i({\bf r},t)+\sum_{i\neq j}G_{ij}({\bf r},t),
\end{equation}
where the $i$-th segment intensity is
\begin{equation}
I_i({\bf r},t)=\frac{1}{2}\epsilon_0c\langle\hat{\bf E}_i^{(-)}({\bf r},t)\hat{\bf E}_i^{(+)}({\bf r},t)\rangle,
\end{equation}
and the correlation function is
\begin{equation}
G_{ij}({\bf r},t)=\frac{1}{2}\epsilon_0c\langle\hat{\bf E}_i^{(-)}({\bf r},t)\mbox{\boldmath$\cdot$}\hat{\bf E}_j^{(+)}({\bf r},t)\rangle.
\end{equation}
We get
\begin{eqnarray}
I_i({\bf r},t)&=&\frac{\mu^2a \omega_i^4}{16\pi^2\epsilon_0c^3L_i}\cot^2\left(\frac{\pi a}{2L_i}\right)\frac{\sin^2\phi_i}{|{\bf r}-{\bf R}_i|^2} \nonumber \\
&\times&\langle B_i^{\dagger}(0)B_i(0)\rangle e^{-\Gamma_i(t-t_i)},
\end{eqnarray}
and
\begin{eqnarray}
G_{ij}({\bf r},t)&=&\frac{\mu^2a \omega_i^2\omega_j^2}{16\pi^2\epsilon_0c^3\sqrt{L_iL_j}}\cot\left(\frac{\pi a}{2L_i}\right)\cot\left(\frac{\pi a}{2L_j}\right)  \nonumber \\
&\times&e^{-\Gamma_i(t-t_i)/2}e^{-\Gamma_j(t-t_j)/2} e^{i\omega_i(t-t_i)}e^{-i\omega_j(t-t_j)}  \nonumber \\
&\times& \langle B_i^{\dagger}(0)B_j(0)\rangle\frac{\sin\phi_i}{|{\bf r}-{\bf R}_i|}\frac{\sin\phi_j}{|{\bf r}-{\bf R}_j|}\left(\hat{\bf n}_i\mbox{\boldmath$\cdot$}\hat{\bf n}_j\right), \nonumber \\
\end{eqnarray}
where we used $E_i=\hbar\omega_i$ and defined $t_i=|{\bf r}-{\bf R}_i|/c$. The results can be easily generalized to the case of finite optical lattice segments that distributed in a plane. In the next section we present the explicit results for a simple case of two optical lattice segments.

\section{Two Finite Optical Lattices}

Let us explicitly look out two finite one dimensional optical lattice segments, $(\alpha)$ and $(\beta)$. The atoms in the two lattices are identical where the only difference is in the length of the lattices. Collective electronic excitations can transfer efficiently among two lattices via collective dipole-dipole interactions only if the two states at the two lattices are in resonance. The transfer is possible for two identical lattices even though they can have different collective transition dipole directions. The transfer energy of a collective electronic excitation among two lattices when their centers are separated by a distance ${\bf R}$ with collective transition dipoles of $\mbox{\boldmath$\mu$}_{\alpha}$ and $\mbox{\boldmath$\mu$}_{\beta}$, as seen in figure (8), is given by
\begin{equation}
J_{\alpha\beta}({\bf R})=\frac{1}{4\pi\epsilon_0}\left\{\frac{(\mbox{\boldmath$\mu$}_{\alpha}\cdot\mbox{\boldmath$\mu$}_{\beta})}{|{\bf R}|^3}-\frac{3(\mbox{\boldmath$\mu$}_{\alpha}\cdot{\bf R})(\mbox{\boldmath$\mu$}_{\beta}\cdot{\bf R})}{|{\bf R}|^5}\right\}.
\end{equation}
If the two lattices are not along the same line and located in a plane, this result holds only when the distance between the lattices is larger than their lengths, that is $|{\bf R}|>L_{\alpha,\beta}$. For example, if we have a lattice with a single site, and a second lattice of two sites. In the lattice of two sites, which are denoted by $(1)$ and $(2)$, we get two collective electronic modes, one mode is dark with transition energy of $E_d=E_A-J_{12}$ and zero transition dipole $\mbox{\boldmath$\mu$}_d=0$, the other mode is bright with transition energy of $E_b=E_A+J_{12}$ and transition dipole of $\mbox{\boldmath$\mu$}_b=\sqrt{2}\mbox{\boldmath$\mu$}$, see figure (9). Here $J_{12}$ is the transfer parameter among the two atoms in the lattice of two sites which is given in (\ref{Coupling}). Note that the parameter $J_{12}$ can be controlled by changing the angle $\theta$. Finally we get in lattice $(\alpha)$ of a single site: $E_{\alpha}=E_A$ and $\mbox{\boldmath$\mu$}_{\alpha}=\mbox{\boldmath$\mu$}$, and in the lattice $(\beta)$ of the two sites: $E_{\beta}=E_A+J_{12}$ and $\mbox{\boldmath$\mu$}_{\beta}=\sqrt{2}\mbox{\boldmath$\mu$}$. In general the energies $E_{\alpha}$ and $E_{\beta}$ are off resonance and the energy transfer is impossible. The energy transfer is possible only if $J_{12}\approx 0$ which is the case at $\theta\approx 54.7$.

\begin{figure}[h!]
\centerline{\epsfxsize=5cm \epsfbox{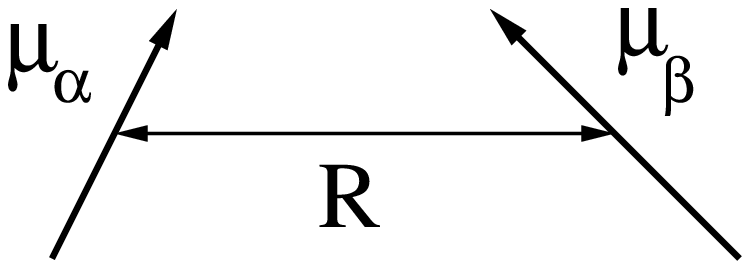}}
\caption{Two transition dipoles $\mbox{\boldmath$\mu$}_{\alpha}$ and $\mbox{\boldmath$\mu$}_{\beta}$ are separated by a distance $|{\bf R}|$.}
\end{figure}

We consider here in details the case of a finite one dimensional optical lattice with a single vacancy that separate the lattice into two segments, $(\alpha)$ and $(\beta)$ of lengths $L_{\alpha}$ and $L_{\beta}$. We consider only the superradiant mode at each segment with energies $E_{\alpha}$ and $E_{\beta}$, and damping rates $\Gamma_{\alpha}$ and $\Gamma_{\beta}$. The collective transition dipoles are taken to be localized at the center of each segment, that is at ${\bf R}_{\alpha}$ and ${\bf R}_{\beta}$. The collective electronic excitation are excited initially, e.g., by a scattering of photons. We consider different cases for the collective electronic excitation states. If we consider a single collective excitation at most in each segment, we have the initial general state
\begin{equation}
|\psi\rangle=C_{00}|0_{\alpha},0_{\beta}\rangle+C_{10}|1_{\alpha},0_{\beta}\rangle+C_{01}|0_{\alpha},1_{\beta}\rangle+C_{11}|1_{\alpha},1_{\beta}\rangle.
\end{equation}
We neglect the first term where no excitations appear in the system. For the initial state $|\psi\rangle=|1_\alpha,0_{\beta}\rangle$ we get $I({\bf r},t)=I_{\alpha}({\bf r},t)$, for the initial state $|\psi\rangle=|0_\alpha,1_{\beta}\rangle$ we get $I({\bf r},t)=I_{\beta}({\bf r},t)$, and for the initial state $|\psi\rangle=|1_\alpha,1_{\beta}\rangle$ we get $I({\bf r},t)=I_{\alpha}({\bf r},t)+I_{\beta}({\bf r},t)$. In all these initial states we have no interference at the observation point. The interference appears for initial states with coherent superposition between the collective electronic excitations. For example, with the initial state
\begin{equation}
|\psi\rangle=\frac{1}{\sqrt{2}}\left(|1_{\alpha},0_{\beta}\rangle+|0_{\alpha},1_{\beta}\rangle\right),
\end{equation}
we get
\begin{equation}
I({\bf r},t)=\frac{1}{2}\left(I_{\alpha}({\bf r},t)+I_{\beta}({\bf r},t)+G_{\alpha\beta}({\bf r},t)+G_{\beta\alpha}({\bf r},t)\right),
\end{equation}
where
\begin{eqnarray}
G_{\alpha\beta}+G_{\beta\alpha}&=&\frac{\mu^2a \omega_{\alpha}^2\omega_{\beta}^2}{8\pi^2\epsilon_0c^3\sqrt{L_{\alpha}L_{\beta}}}\cot\left(\frac{\pi a}{2L_{\alpha}}\right)\cot\left(\frac{\pi a}{2L_{\beta}}\right) \nonumber \\
&\times&\frac{\sin\phi_{\alpha}}{|{\bf r}-{\bf R}_{\alpha}|}\frac{\sin\phi_{\beta}}{|{\bf r}-{\bf R}_{\beta}|} \left(\hat{\bf n}_{\alpha}\mbox{\boldmath$\cdot$}\hat{\bf n}_{\beta}\right)e^{-\Gamma_{\alpha}(t-t_{\alpha})/2} \nonumber \\
&\times&e^{-\Gamma_{\beta}(t-t_{\beta})/2}\cos\left[\omega_{\alpha}(t-t_{\alpha})-\omega_{\beta}(t-t_{\beta})\right], \nonumber \\
\end{eqnarray}
and for $(i=\alpha,\beta)$ we have
\begin{equation}
I_i({\bf r},t)=\frac{\mu^2a \omega_i^4}{16\pi^2\epsilon_0c^3L_i}\cot^2\left(\frac{\pi a}{2L_i}\right)\frac{\sin^2\phi_i}{|{\bf r}-{\bf R}_i|^2} e^{-\Gamma_i(t-t_i)}.
\end{equation}

\begin{figure}[h!]
\centerline{\epsfxsize=8cm \epsfbox{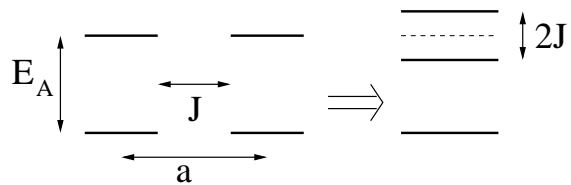}}
\caption{In the left, two two-level atoms of transition energy $E_A$ are separated by a distance $a$ with energy transfer parameter $J$. In the right, the two collective electronic excitation states, the symmetric and antisymmetric states, of energies $E_A+J$ and $E_A-J$ and with splitting energy of $2J$.}
\end{figure}

Here we concentrate in a simple example of two lattice segments one includes a single site and the other includes two sites, along the $\hat{\bf x}$ axis, as appears in figure (10). Segment $(\alpha)$, of length $L_{\alpha}=2a$, is located at the origin ${\bf R}_{\alpha}=(0,0,0)$ with an atom of transition energy $E_{\alpha}=\hbar\omega_A$ and transition dipole $\mbox{\boldmath$\mu$}_{\alpha}=\mbox{\boldmath$\mu$}$ where $\mbox{\boldmath$\mu$}=\mu(\cos\theta,0,\sin\theta)$. The excited state damping rate is $\Gamma_{\alpha}=\Gamma_A$. Segment $(\beta)$, of length $L_{\beta}=3a$, includes an atom at the position $(2a,0,0)$ and a second at $(3a,0,0)$ with the same transition energy and dipole as in segment $(\alpha)$. The vacancy is at position $(a,0,0)$. The resonance dipole-dipole interaction induces two collective electronic excitation modes, as discussed previously. The antisymmetric and dark mode of energy $E_a=E_A-J$ and zero transition dipole, and the symmetric and bright mode of energy $E_s=E_A+J$ and transition dipole $\sqrt{2}\mbox{\boldmath$\mu$}$. We consider for segment $(\beta)$ only the bright state which is denoted as before by $E_{\beta}=E_A+J$ and $\mbox{\boldmath$\mu$}_{\beta}=\sqrt{2}\mbox{\boldmath$\mu$}$. The bright state damping rate is $\Gamma_{\beta}=2\Gamma_A$. For segment $(\alpha)$ the transition dipole is localized at ${\bf R}_{\alpha}=(0,0,0)$, and for segment $(\beta)$, as the observation point is in the far zone, the collective transition dipole is taken to be at ${\bf R}_{\beta}=(\frac{5}{2}a,0,0)$. We define also the distance vector between the centers of the two segments by $\bar{\bf R}={\bf R}_{\beta}-{\bf R}_{\alpha}$, where here $\bar{R}=\frac{5}{2}a$.

\begin{figure}[h!]
\centerline{\epsfxsize=7cm \epsfbox{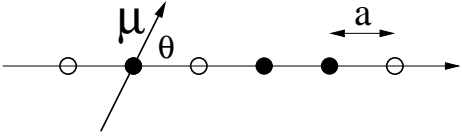}}
\caption{A lattice of two segments. One segment includes a single site, and the other is of two sites. The segments are separated by a single empty site.}
\end{figure}

The observation point is located along the $\hat{\bf z}$ axis at the point ${\bf r}=(0,0,r)$, see figure (12). The distance vector of the observation point from segment $(\alpha)$ is ${\bf r}-{\bf R}_{\alpha}={\bf r}$ which are separated by a distance $|{\bf r}-{\bf R}_{\alpha}|=r$. Segment $(\beta)$ is separated from the observation point by the vector ${\bf r}-{\bf R}_{\beta}=(\bar{R},0,r)$ of distance $|{\bf r}-{\bf R}_{\beta}|=\sqrt{r^2+\bar{R}^2}$. We have the unit vectors $\hat{\bf n}_{\alpha}=\frac{1}{r}(0,0,r)$ and $\hat{\bf n}_{\beta}=\frac{1}{\sqrt{r^2+\bar{R}^2}}(\frac{5}{2}a,0,r)$, where we get $\left(\hat{\bf n}_{\alpha}\mbox{\boldmath$\cdot$}\hat{\bf n}_{\beta}\right)=\frac{r}{\sqrt{r^2+\bar{R}^2}}$. The angle between the vector ${\bf r}-{\bf R}_{\alpha}$ and the transition dipole $\mbox{\boldmath$\mu$}_{\alpha}$ is $\phi_{\alpha}=\frac{\pi}{2}-\theta$, and the angle between ${\bf r}-{\bf R}_{\beta}$ and $\mbox{\boldmath$\mu$}_{\beta}$ is $\phi_{\beta}=\pi-\theta-\varphi$ where $\tan\varphi=\frac{r}{\bar{R}}$, as plotted in figure (11). Collecting all these data we get for the intensities
\begin{equation}
I_{\alpha}({\bf r},t)=\frac{\mu^2\omega_A^4}{32\pi^2\epsilon_0c^3}\ \frac{\sin^2\phi_{\alpha}}{r^2}\ e^{-\Gamma_A\left(t-\frac{r}{c}\right)},
\end{equation}
and
\begin{equation}
I_{\beta}({\bf r},t)=\frac{\mu^2(\omega_A+\bar{J})^4}{16\pi^2\epsilon_0c^3}\ \frac{\sin^2\phi_{\beta}}{r^2+\bar{R}^2}\ e^{-2\Gamma_A\left(t-\frac{1}{c}\sqrt{r^2+\bar{R}^2}\right)},
\end{equation}
with the correlation term
\begin{eqnarray}
&&G_{\alpha\beta}+G_{\beta\alpha}=\frac{\mu^2 \omega_A^2(\omega_A+\bar{J})^2}{8\sqrt{2}\pi^2\epsilon_0c^3}\ \frac{\sin\phi_{\alpha}\sin\phi_{\beta}}{r^2+\bar{R}^2} \nonumber \\
&\times&e^{-\frac{\Gamma_A}{2}\left(t-\frac{r}{c}\right)}e^{-\Gamma_A\left(t-\frac{1}{c}\sqrt{r^2+\bar{R}^2}\right)} \nonumber \\
&\times&\cos\left[\omega_A\left(t-\frac{r}{c}\right)-(\omega_A+\bar{J})\left(t-\frac{1}{c}\sqrt{r^2+\bar{R}^2}\right)\right], \nonumber \\
\end{eqnarray}
where we defined $J=\hbar\bar{J}$. In the far zone limit, where $r\gg \bar{R}$, we have $\phi_{\alpha}\approx\phi_{\beta}=\phi$, we can use
\begin{equation}
I_{\alpha}({\bf r},t)\simeq\frac{\mu^2\omega_A^4}{32\pi^2\epsilon_0c^3}\ \frac{\sin^2\phi}{r^2}\ e^{-\Gamma_A\left(t-\frac{r}{c}\right)},
\end{equation}
and
\begin{equation}
I_{\beta}({\bf r},t)\simeq\frac{\mu^2(\omega_A+\bar{J})^4}{16\pi^2\epsilon_0c^3}\ \frac{\sin^2\phi}{r^2}\ e^{-2\Gamma_A\left(t-\frac{r}{c}\right)},
\end{equation}
with
\begin{eqnarray}
G_{\alpha\beta}+G_{\beta\alpha}&\simeq&\frac{\mu^2 \omega_A^2(\omega_A+\bar{J})^2}{8\sqrt{2}\pi^2\epsilon_0c^3}\ \frac{\sin^2\phi}{r^2}\ e^{-\frac{3\Gamma_A}{2}\left(t-\frac{r}{c}\right)} \nonumber \\
&\times&\cos\left[(\omega_A+\bar{J})\frac{\bar{R}^2}{2rc}-\bar{J}\left(t-\frac{r}{c}\right)\right].
\end{eqnarray}

\begin{figure}[h!]
\centerline{\epsfxsize=6cm \epsfbox{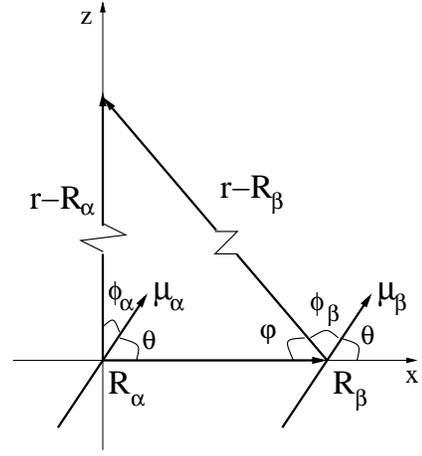}}
\caption{The two collective transition dipoles, $\mbox{\boldmath$\mu$}_{\alpha}$ and $\mbox{\boldmath$\mu$}_{\beta}$, are localized at ${\bf R}_{\alpha}$ and ${\bf R}_{\beta}$, with the observation point at ${\bf r}$. The angle between the vectors ${\bf r}-{\bf R}_{\alpha}$ and $\mbox{\boldmath$\mu$}_{\alpha}$ is $\phi_{\alpha}$, and between the vectors ${\bf r}-{\bf R}_{\beta}$ and $\mbox{\boldmath$\mu$}_{\beta}$ is $\phi_{\beta}$.}
\end{figure}

In figure (12) we plot the scaled intensity $I({\bf r},t)/I_0$ as a function of time $t$, where $I_0=I(t=r/c)$. We used the previous numbers. The observation point is at ${\bf r}=100a\ \hat{\bf z}$. The oscillations of the intensity at the observation point is the signature for the formation of collective electronic excitations among the atoms in the segment of two sites. Also these quantum beats can be used to detect the properties of the energy transfer parameters. Furthermore, the result can be used to fix the relative position of the segments by observing the intensity at a given time for different points in the space.

\begin{figure}[h!]
\centerline{\epsfxsize=8cm \epsfbox{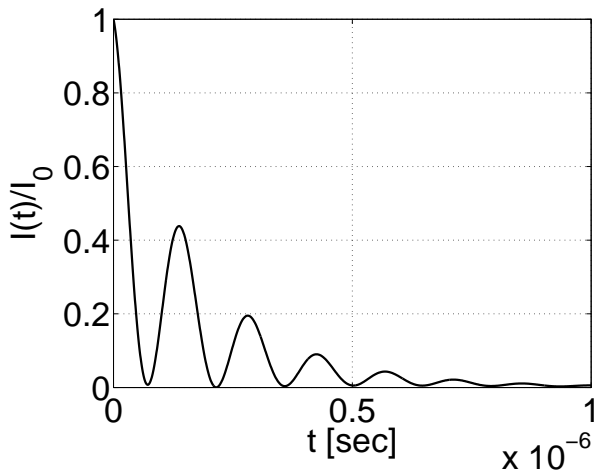}}
\caption{The emission intensity $I({\bf r},t)$ relative to the initial intensity $I_0=I(t=r/c)$ as a function of time, that is $I({\bf r},t)/I_0$ vs. $t$, at the observation point $r=100a$.}
\end{figure}

\section{Summary}

We investigated a finite one dimensional optical lattice with one atom per site in exploiting the collective electronic excitation picture. Here the collective states are standing waves of symmetric and antisymmetric modes. The calculated transition dipole, damping rate and emitted light of the collective modes show that the antisymmetric modes are dark and the symmetric ones are bright. Furthermore, the first bright mode, the one without nodes, found to be superradiant and dominates the optical properties of the finite lattice. Next, we treated finite one dimensional optical lattice with defects, where several vacancies appear at different lattice sites. These vacancies divide the optical lattice into smaller finite optical lattice segments of different lengths. Each segment is treated as in the first part, and the emission pattern of the whole system is obtained by summing over the contributions of all segments.

We emphasized the case of two optical lattice segments of different number of sites and that separated by a single empty site. Interactions of collective excitations at the two segments are discussed, and we show how energy transfer among segments of different lengths can be blocked due to the fact that the collective excitations have different energies and are off resonance. We presented the results for a simple case of two segments, one with a single site and the other with two sites. The emitted light can provide us with the physical properties of the system concerning the formation of collective excitations and resonance dipole-dipole interactions, beside the relative position of the optical lattice segments.

The discussion in the present paper can be easily generalized to treat finite optical lattice segments, of different lengths and directions, which are distributed in the plane of two dimensional optical lattice. More complex patterns of different shapes, which build of linear atomic chains in a plane, can be considered by numerically calculating their collective excitations. For optical lattices with small lattice constant or strong transition dipole the formation of collective excitations is unavoidable, and the considerations of the present paper are desirable.

\ 

The work was supported by the Austrian Science Funds (FWF), via the project (P21101).



\begin{thebibliography}{}

\bibitem{Metcalf} H. J. Metcalf, and P. van der Straten, {\it Laser Cooling and Trapping}, (Springer, NY, 1999).

\bibitem{Zeilinger} D. Bouwmeester, A. K. Ekert, and A. Zeilinger, {\it The physics of Quantum Information}, (Springer, NY, 2000).

\bibitem{Dalibard} I. Bloch, J. Dalibard, and W. Zwerger, {\it Rev. Mod. Phys.} {\bf 80}, 885 (2008).

\bibitem{Lewenstein} M. Lewenstein, A. Sanpera, V. Ahufinger, B. Damski, A. Sen De, and U. Sen, {\it Adv. in Phys.} {\bf 56}, 243 (2007).

\bibitem{Jaksch} D. Jaksch, C. Bruder, J. I. Cirac, C. W. Gardiner, and P. Zoller, {\it Phys. Rev. Lett.} {\bf 81}, 3108 (1998).

\bibitem{Bloch} M. Greiner, O. Mandel, T. Esslinger, T. W. Hansch, and I. Bloch, {\it Nature} {\bf 415}, 39 (2002).

\bibitem{Spielman} I. B. Spielman, W. D. Phillips, and J. V. Porto, {\it Phys. Rev. Lett.} {\bf 98}, 080404 (2007).

\bibitem{Rauschenbeutel} G. Sague, E. Vetsch, W. Alt, D. Meschede, and A. Rauschenbeutel, {\it Phys. Rev. Lett.} {\bf 99}, 163602 (2007).

\bibitem{Nayak} K. P. Nayak, P. N. Melentiev, M. Morinaga, F. L. Kien, V. I. Balykin, and K. Hakuta, {\it Opt. Express} {\bf 15}, 5431 (2007).

\bibitem{Vetsch} E. Vetsch, D. Reitz, G. Sague, R. Schmidt, S. T. Dawkins, and A. Rauschenbeutel, {\it Phys. Rev. Lett.} {\bf 104}, 203603 (2010).

\bibitem{Sherson} J. F. Sherson, C. Weitenberg, M. Endres, M. Cheneau, I. Bloch, and S. Kuhr, {\it Nature} {\bf 467}, 68 (2010).

\bibitem{Bakr} W. S. Bakr, A. Peng, M. E. Tai, R. Ma, J. Simon, J. I. Gillen, S. Folling, L. Pollet, and M. Greiner, {\it Science} {\bf 329}, 547 (2010).

\bibitem{Weitenberg} C. Weitenberg, M. Endres, J. F. Sherson, M. Cheneau, P. Schauss, T. Fukuhara, I. Bloch, and S. Kuhr, {\it Nature} {\bf 471}, 319 (2011).

\bibitem{ZoubiA} H. Zoubi, and H. Ritsch, {\it Phys. Rev. A} {\bf 76}, 013817 (2007).

\bibitem{Davydov} S. Davydov, {\it Theory of Molecular Excitons}, (Plenum, New York, 1971).

\bibitem{Agranovich} V. M. Agranovich, {\it Excitations in Organic Solids}, (Oxford, UK, 2009).

\bibitem{ZoubiB} H. Zoubi, and H. Ritsch, {\it Europhys. Lett.} {\bf 87}, 23001 (2009).

\bibitem{ZoubiC} H. Zoubi, and H. Ritsch, {\it Europhys. Lett.} {\bf 90}, 23001 (2010).

\bibitem{ZoubiD} H. Zoubi, and H. Ritsch, arXiv:1103.2949. {\it Phys. Rev. A} (in press).

\bibitem{ZoubiE} H. Zoubi, and H. Ritsch, {\it New J. Phys.} {\bf 10}, 23001 (2008).

\bibitem{ZoubiF} H. Zoubi, and H. Ritsch, {\it New J. Phys.} {\bf 12}, 103014 (2010).

\bibitem{Loudon} R. Loudon, {\it The Quantum Theory of Light}, 3rd Ed. (Oxford, UK, 2000).

\end{thebibliography}
\end{document}